\documentclass[pra,twocolumn,showpacs]{revtex4}
\usepackage{graphics}       
\usepackage{bm}         
\usepackage{amsmath}        
\usepackage{latexsym}
\usepackage{amsfonts}
\usepackage{amssymb}
\begin{document}
\title{Condition for unambiguous state discrimination using local operations and classical communication}
\author{Anthony Chefles}
\affiliation{School of Physics, Astronomy and Mathematics,
University of Hertfordshire,
       Hatfield AL10 9AB, Hertfordshire, UK}

\begin{abstract}
\vspace{0.5cm} We obtain a necessary and sufficient condition for
a finite set of states of a finite dimensional multiparticle
quantum system to be amenable to unambiguous discrimination using
local operations and classical communication. This condition is
valid for states which may be be mixed, entangled or both. When
the support of the set of states is the entire multiparticle
Hilbert space, this condition is found to have an intriguing
connection with the theory of entanglement witnesses.

\end{abstract}
 \maketitle

It was discovered by Walgate et al \cite{Walgate1} that one can perfectly distinguish between any
pair of orthogonal, bipartite, pure entangled states without collective measurements. It suffices
to use only local operations and classical communication (LOCC) between the parties who possess
the component systems.

This discovery runs counter to some of our intuitions about
entangled states.  Part or all of the information in an entangled
state will be stored in Einstein-Podolsky-Rosen (EPR)-type
correlations \cite{EPR} between the subsystems rather than in
properties of the subsystems themselves. It would therefore be
reasonable to expect that perfect discrimination among locally
similar orthogonal states, which differ mainly in terms of their
collective, EPR correlations, would require a collective
measurement.

The discovery that this is not always the case has given rise to a
considerable amount of activity devoted to exploring this
phenomenon. Particular questions which have been raised include
the following['[': under what circumstances is perfect LOCC
discrimination among orthogonal states possible?  This relates to
the `nonlocality without entanglement' discovered by Bennett et al
\cite{NWE} and numerous interesting results relating to this
problem have been obtained by several authors
\cite{Unlockable,Ghosh1,Walgate2,Ghosh2,Chen2,Chen3}. An
intriguing, persistent theme here has been the relationship
between perfect LOCC state discrimination and entanglement
distillation.

Another issue that has been the focus of much attention has been LOCC discrimination among a set
of non-orthogonal states. If the states are non-orthogonal, then perfect discrimination among
them is impossible and we must settle for an imperfect strategy. One such strategy is minimum
error discrimination. Virmani et al \cite{Virmani1} showed that any two multiparticle pure states
can be discriminated with minimum error probability using LOCC operations. An alternative
discrimination strategy is unambiguous state discrimination, where the result is never incorrect
though it may be inconclusive. These authors also considered the possibility that this set of
states can be unambiguously discriminated with the minimum probability of inconclusive results by
an LOCC measurement.  This was subsequently proven to be true for a pair of equally-probable pure
states by Chen and Yang \cite{Chen1}. More recently, Hillery and Mimih \cite{Hillery} have
investigated LOCC unambiguous state discrimination with certain restrictions placed on the
permitted types of classical communication.  Also, Horodecki et al \cite{Horodecki} have obtained
a necessary and sufficient condition for LOCC unambiguous discrimination among a complete set of
orthogonal pure states.

All of these results beg the question: which sets of states can be
unambiguously discriminated using an LOCC procedure? Even without
the LOCC constraint, the linearity of quantum mechanics imposes
restrictions on the kind of state sets for which unambiguous
discrimination is possible. In particular, if the possible states
are pure, and all states have non-zero detection probability, then
they must be linearly independent \cite{Linear}. We anticipate
that, in general, the conditions which determine when a set of
multiparticle states can be unambiguously discriminated with LOCC
will be stronger than this and will relate in some way to the
local nature of the measurement. Our search for such conditions
could be impeded by the fact that LOCC measurements are
notoriously difficult to characterise.  At the time of writing,
the structure of the set of LOCC measurements is effectively
unknown.  However, such measurements form a subset of the much
simpler and completely characterised set of separable
measurements.   The main result of this paper is a theorem which
exploits the simple form of separable measurements to derive a
necessary condition for separable and therefore LOCC unambiguous
state discrimination for any finite number of quantum states of an
arbitrary, finite dimensional, multiparticle quantum system. This
condition in then proven, in a constructive way, to be sufficient
for LOCC, and therefore also separable unambiguous state
discrimination under the specified conditions. Thus, the necessary
and sufficient conditions for LOCC unambiguous state
discrimination are identical to those for the broader and simpler
class of separable measurements.  Probabilistic LOCC
implementation of separable quantum operations has also been
considered in \cite{Cirac}.

The remainder of this paper is devoted to exploring some
implications of this result.  We find it to be equivalent to the
condition for unambiguous state discrimination without the LOCC
constraint supplemented by a subsidiary condition which exhibits
the local nature of the measurement. It is also shown that there
are curious links between LOCC unambiguous state discrimination
and entanglement witnesses.

Prior to discussing the problem of LOCC unambiguous state discrimination, we shall find it to be
helpful to give a reformulation of the conditions for unambiguous state discrimination not
subject to the LOCC constraint.  In the context of this paper, it is appropriate to refer to this
as unconstrained unambiguous state discrimination.

The situation we consider here is as follows: we have a quantum
system $q$ with associated Hilbert space ${\cal H}$, which we take
to have finite dimensionality $D$.  The system is prepared in one
of $M<{\infty}$ quantum states ${\rho}_{\mu}$, where
${\mu}{\in}\{1,{\ldots},M\}$.  We would like to determine
unambiguously which state has been prepared.

It may be the case that we are only interested in obtaining a conclusive identification, with
non-zero probability, for a subset of the $M$ possible states.  So, let ${\Delta}$ be the subset
of $\{1,{\ldots},M\}$ for which the discrimination probability is required to be greater than
zero for ${\rho}_{\mu}$ and ${\mu}{\in}{\Delta}$.  Also, let $M_{\Delta}$ be the number of
elements of the set ${\Delta}$.

We require a measurement with $M_{\Delta}+1$ outcomes: $M_{\Delta}$ of these correspond to the
states which are to be positively discriminated and a further outcome signals an inconclusive
result, `?'.  These requirements can be expressed in terms of positive, operator-valued measures
(POVMs). We need $M_{\Delta}$ positive operators ${\Pi}_{{\mu}}$ and a further positive operator
${\Pi}_{?}$ such that
\begin{equation}
{\Pi}_{?}+\sum_{{\mu}{\in}{\Delta}}{\Pi}_{\mu}=1_{\cal H},
\end{equation}
where $1_{\cal H}$ is the identity operator on ${\cal H}$.  The
POVM elements satisfy
\begin{equation}
\mathrm{Tr}({\rho}_{{\mu}}{\Pi}_{{\mu}'})=p({\mu}|{\rho}_{\mu}){\delta}_{{\mu}{\mu}'}.
\end{equation}
Here, $p({\mu}|{\rho}_{\mu})>0\;\;{\forall}\;\;{\mu}{\in}{\Delta}$
is the probability of detecting the state ${\rho}_{{\mu}}$.

 In
order to obtain the conditions under which such a measurement is possible, we make the following
definitions.  Let ${\cal R}_{\mu}$ be the support \cite{support} of ${\rho}_{\mu}$ and let ${\cal
R}_{\bar{\mu}}$ be the support of $(\sum_{{\mu}'{\neq}{\mu}=1}^{M}{\rho}_{{\mu}'})$, that is, the
sum of the $M-1$ other density operators. Let us denote the orthogonal complements of ${\cal
R}_{\mu}$ and ${\cal R}_{\bar{\mu}}$ by ${\cal R}^{\perp}_{\mu}$ and ${\cal
R}^{\perp}_{\bar{\mu}}$ respectively.  We have ${\cal R}_{\mu}{\oplus}{\cal
R}^{\perp}_{\mu}={\cal R}_{\bar{\mu}}{\oplus}{\cal R}^{\perp}_{\bar{\mu}}={\cal H}$. Furthermore,
let us denote the set complement of ${\cal R}^{\perp}_{\mu}$ by $\bar{\cal R}^{\perp}_{\mu}$.
That is, ${\cal R}^{\perp}_{\mu}{\cup}\bar{\cal R}^{\perp}_{\mu}={\cal H}$ and ${\cal
R}^{\perp}_{\mu}{\cap}\bar{\cal R}^{\perp}_{\mu}={\emptyset}$.  Here, ${\emptyset}$ is the empty
set. Finally, let ${\cal S}_{\mu}={\cal R}^{\perp}_{\bar{\mu}}{\cap}\bar{\cal R}^{\perp}_{\mu}$.

A necessary and sufficient condition for unconstrained unambiguous
discrimination among the $M$ states ${\rho}_{\mu}$, with non-zero
detection probability for ${\mu}{\in}{\Delta}$, is that for all
${\mu}{\in}{\Delta}$,
\begin{equation}
{\cal S}_{\mu}{\neq}{\emptyset}.
\end{equation}
To prove necessity, we see from Eq. (2) that for each
${\mu}'{\in}{\Delta}$, ${\Pi}_{{\mu}'}$ has an eigenstate,
$|{\pi}_{{\mu}'}{\rangle}$, such that
\begin{equation}
{\langle}{\pi}_{{\mu}'}|{\rho}_{\mu}|{\pi}_{{\mu}'}{\rangle}>0\;\;\mathrm{iff}\;\;{\mu}={\mu}',
\end{equation}
for all ${\mu}{\in}\{1,{\ldots},M\}$.  This implies that, for ${\mu}{\in}{\Delta}$,
$|{\pi}_{\mu}{\rangle}$ is in both the sets ${\cal R}^{\perp}_{\bar{\mu}}$ and $\bar{\cal
R}^{\perp}_{\mu}$. It is thus in their intersection, and thus ${\cal S}_{\mu}$. This implies that
${\cal S}_{\mu}$ cannot be the empty set or the set whose sole element is the zero vector.
However, by definition, ${\cal S}_{\mu}$ cannot contain the zero vector.  The reason for this is
that the set ${\cal R}^{\perp}_{\mu}$, being a vector space, contains the zero vector and so its
set complement $\bar{\cal R}^{\perp}_{\mu}$ doesn't, and every element of ${\cal S}_{\mu}$ is, by
definition, also an element of $\bar{\cal R}^{\perp}_{\mu}$.  So, we see that  ${\cal S}_{\mu}$
cannot be the empty set and this proves necessity.

To prove sufficiency, we note that for each ${\mu}{\in}{\Delta}$,
if ${\cal S}_{\mu}$ is non-empty then it must contain at least one
pure state.  Let $|{\pi}_{\mu}{\rangle}$ be one pure state in
${\cal S}_{\mu}$. This state is clearly also an element of both
${\cal R}^{\perp}_{\bar{\mu}}$ and $\bar{\cal R}^{\perp}_{\mu}$,
since ${\cal S}_{\mu}$ is the intersection of these sets.

Let ${\lambda}$ be the maximum eigenvalue of the operator
$\sum_{{\mu}{\in}{\Delta}}|{\pi}_{\mu}{\rangle}{\langle}{\pi}_{\mu}|$. The fact that this
operator acts on a finite dimensional Hilbert space implies that it is bounded \cite{Kreyzig},
and so ${\lambda}<{\infty}$. Consider now the POVM:
\begin{eqnarray}
{\Pi}_{\mu}&=&{\lambda}^{-1}|{\pi}_{\mu}{\rangle}{\langle}{\pi}_{\mu}|\;\;\mathrm{for}\;\;{\mu}{\in}{\Delta}
\\ {\Pi}_{?}&=&1_{\cal H}-\sum_{{\mu}{\in}{\Delta}}{\Pi}_{\mu}.
\end{eqnarray}
Then we have
\begin{equation}
\mathrm{Tr}({\rho}_{\mu}{\Pi}_{{\mu}'})=p({\mu}|{\rho}_{\mu}){\delta}_{{\mu}{\mu}'},\;\;\;\;p({\mu}|{\rho}_{\mu})>0\;\;\mathrm{for}\;\;{\mu}{\in}{\Delta},
\end{equation}
where
$p({\mu}|{\rho}_{\mu})={\lambda}^{-1}{\langle}{\pi}_{\mu}|{\rho}_{\mu}|{\pi}_{\mu}{\rangle}$.
We know from Eq. (2) that
${\langle}{\pi}_{\mu}|{\rho}_{{\mu}}|{\pi}_{\mu}{\rangle}$ is
non-zero.  Also, the fact that ${\lambda}<{\infty}$ implies that
${\lambda}^{-1}>0$. Taken together, these observations lead to the
conclusion that
$p({\mu}|{\rho}_{\mu})>0\;\;{\forall}\;\;{\mu}{\in}{\Delta}$ and
this proves sufficiency.

Let us now bring these ideas to bear on the problem of LOCC
unambiguous discrimination among $M<{\infty}$ multiparticle
quantum states ${\rho}_{\mu}$. Here, we have $N$ quantum systems
$q_{j}$, where $j{\in}\{1,{\ldots},N\}$. System $q_{j}$ has
associated Hilbert space ${\cal H}_{j}$, having finite
dimensionality $D_{j}$. The Hilbert space of the composite system
is the tensor product space ${\cal H}={\otimes}_{j=1}^{N}{\cal
H}_{j}$ and we shall denote the dimensionality of this space by
$D={\prod}_{j=1}^{N}D_{j}$.  Each of the $N$ systems is spatially
separated from the others.

The states to be discriminated, ${\rho}_{\mu}$, are states of this $N$ particle system.  The $N$
parties who each possess one of the component subsystems are restricted to the use of local
operations in their individual laboratories and classical communication between them in order to
collectively unambiguously determine the state.  This LOCC constraint will impose restrictions
upon the form of the POVM elements ${\Pi}_{\mu}$.  While general, simple necessary and sufficient
criteria for a POVM to be implementable by LOCC are not known at this time, one useful necessary
condition \cite{NWE} is that the POVM elements must be separable. This means that they have to be
of the form
\begin{equation}
{\Pi}_{{\mu}'}=\sum_{k=1}^{K}{\otimes}_{j=1}^{N}{\Pi}_{{\mu}'jk}
\end{equation}
for some $K$, and where the positive operator ${\Pi}_{{\mu}'jk}$ acts upon the space ${\cal
H}_{j}$.  As before, ${\mu}'{\in}{\Delta}$.  We may, without loss of generality, take
$K{\leq}D^{2}$, which is finite. We will now use this expression to derive a necessary condition
for LOCC unambiguous state discrimination which we will also constructively prove to be
sufficient.\\

\noindent $\mathbf{Theorem}$ {\em A condition for unambiguous
discrimination among $M$ $N$-particle states ${\rho}_{\mu}$, with
non-zero detection probability for ${\mu}{\in}{\Delta}$, which is
necessary for separable (and therefore LOCC) measurements and
sufficient for LOCC (and therefore separable) measurements is that
for each ${\mu}{\in}{\Delta}$, the set ${\cal S}_{\mu}$ contains
an $N$-particle product state.}\\

\noindent $\mathbf{Proof}$.  We will prove necessity first.
Substituting (8) into (2), we see that the requirement of
unambiguous state discrimination can be expressed as
\begin{equation}
\mathrm{Tr}\left({\rho}_{\mu}\sum_{k=1}^{K}{\otimes}_{j=1}^{N}{\Pi}_{{{\mu}'}jk}\right)=p({\mu}|{\rho}_{\mu}){\delta}_{{\mu}{{\mu}'}},
\end{equation}
where $p({\mu}|{\rho}_{\mu})>0$ for all ${\mu}{\in}{\Delta}$. Coupling this with the fact that
$K<{\infty}$ implies that for each ${\mu}'{\in}{\Delta}$, there exists
$k_{{\mu}'}{\in}\{1,{\ldots},K\}$ such that
\begin{equation}
\mathrm{Tr}\left({\rho}_{\mu}{\otimes}_{j=1}^{N}{\Pi}_{{{\mu}'}jk_{{\mu}'}}\right)=p'({\mu}|{\rho}_{\mu}){\delta}_{{\mu}{{\mu}'}}
\end{equation}
for some $p'({\mu}|{\rho}_{\mu})>0$.  The eigenstates of
${\otimes}_{j=1}^{N}{\Pi}_{{{\mu}'}jk_{{\mu}'}}$ may be taken to be of product form.   It follows
from Eq. (10) that for each ${\mu}'{\in}{\Delta}$,
${\otimes}_{j=1}^{N}{\Pi}_{{{\mu}'}jk_{{\mu}'}}$ has a product state eigenstate,
$|{\pi}_{{{\mu}'}}{\rangle}$, such that
\begin{equation}
{\langle}{\pi}_{{\mu}'}|{\rho}_{{\mu}}|{\pi}_{{\mu}'}{\rangle}=\tilde{p}({\mu}|{\rho}_{\mu}){\delta}_{{\mu}{{\mu}'}}
\end{equation}
and $\tilde{p}({\mu}|{\rho}_{\mu})>0$ for all ${\mu}{\in}{\Delta}$. This is a necessary condition
for separable or LOCC unambiguous state discrimination. It is clear from this expression that,
for ${\mu}{\in}{\Delta}$, $|{\pi}_{\mu}{\rangle}$ must be orthogonal to ${\cal R}_{\bar{\mu}}$
and that it is therefore an element of ${\cal R}^{\perp}_{\bar{\mu}}$. The fact that
$\tilde{p}({\mu}|{\rho}_{\mu})>0$ implies that $|{\pi}_{\mu}{\rangle}$ cannot be an element of
${\cal R}^{\perp}_{\mu}$.  It must be an element of its set complement $\bar{\cal
R}^{\perp}_{\mu}$. The product state $|{\pi}_{\mu}{\rangle}$ is an element of both ${\cal
R}^{\perp}_{\bar{\mu}}$ and $\bar{\cal R}^{\perp}_{\mu}$ and so it is also an element of their
intersection ${\cal S}_{\mu}$. This proves the necessity of our condition.

We shall now see that this is also a sufficient condition. Suppose
that, for each ${\mu}{\in}{\Delta}$, ${\cal S}_{\mu}$ contains a
product state $|{\pi}_{{\mu}}{\rangle}$.  These states then
satisfy Eq. (11) and may be written as
$|{\pi}_{\mu}{\rangle}={\otimes}_{j=1}^{N}|{\pi}_{{\mu}j}{\rangle}$.
 Let ${\lambda}_{j}$ be the maximum eigenvalue of the operator $\sum_{{\mu}{\in}{\Delta}}|{\pi}_{{\mu}j}{\rangle}{\langle}{\pi}_{{\mu}j}|$.  Consider now performing a local measurement on the $j$th system
with $M_{\Delta}+1$ outcomes, having the corresponding POVM
elements:
\begin{equation}
{\Pi}_{{\mu}j}={\lambda}_{j}^{-1}|{\pi}_{{\mu}j}{\rangle}{\langle}{\pi}_{{\mu}j}|,\;\;\;\;{\Pi}_{?j}=1_{{\cal
H}_{j}}-\sum_{{\mu}{\in}{\Delta}}{\Pi}_{{\mu}j},
\end{equation}
where $1_{{\cal H}_{j}}$ is the identity operator on ${{\cal
H}_{j}}$.  All of these operators are positive.  Following these
measurements, the results are discussed among the $N$ parties over
classical communication channels.  It is easy to see that if all
parties obtain the same result `${\mu}$', then it follows that the
initial state must have been ${\rho}_{\mu}$.  This will occur with
probability
$p({\mu}|{\rho}_{\mu})=\tilde{p}({\mu}|{\rho}_{\mu}){\prod}_{j=1}^{N}{\lambda}_{j}^{-1}$.
For all ${\mu}{\in}{\Delta}$, this can be seen to be non-zero. The
$\tilde{p}({\mu}|{\rho}_{\mu})$ are non-zero as a consequence of
Eq. (11). Furthermore, the ${\lambda}_{j}^{-1}$ are also non-zero
because the ${\Pi}_{{\mu}j}$, being operators on
finite dimensional spaces, are bounded.  This implies that ${\lambda}_{j}<{\infty}$ and completes the proof.${\Box}$\\

Let us make some observations about this theorem.  Firstly, it is interesting to see how the
above condition for LOCC unambiguous state discrimination is stronger than that for unconstrained
unambiguous state discrimination.  If arbitrary quantum measurements are possible, then we know
from Eq. (3) that the necessary and sufficient condition for unambiguous state discrimination is
that for all ${\mu}{\in}{\Delta}$, ${\cal S}_{\mu}{\neq}{\emptyset}$.  When the states are pure
and ${\Delta}=\{1,{\ldots},M\}$, this is equivalent to the condition that the states are linearly
independent.

However, the above theorem shows that, to perform unambiguous
state discrimination by LOCC means, not only must each of these
sets be non-empty, they must also each contain a product state. We
have thus separated the condition for LOCC unambiguous state
discrimination into the condition for unconstrained unambiguous
state discrimination and a subsidiary condition which exhibits the
local nature of the measurement procedure.

Notice that classical communication barely enters into the discussion. In fact, the measurement
we constructed in our proof of the sufficiency part of the theorem only used classical
communication to establish whether or not the discrimination attempt was a success. In
particular, no local measurement on one system depended on the results of local measurements on
other systems. This shows that if we are only interested in performing LOCC unambiguous state
discrimination with some non-zero probability, and are not interested in optimality, then the
significance of classical communication is limited to the collective establishment of success or
failure. It has however been shown by Hillery and Mimih \cite{Hillery} that if we wish to perform
optimal LOCC unambiguous state discrimination, then classical communication channels play a more
significant role.

Let us now examine in more detail the ${\cal S}_{\mu}$ for ${\mu}{\in}{\Delta}$.  The key
property of presence or absence of product states in these sets suggests a connection with
entanglement witnesses, which discriminate between product and entangled states
\cite{Terhal,Bruss}.  Recall that an Hermitian operator $W$ is an entanglement witness iff, for
every separable state ${\rho}_{sep}$, $\mathrm{Tr}({\rho}_{sep}W){\geq}0$ and for at least one
entangled state ${\rho}_{ent}$, $\mathrm{Tr}({\rho}_{ent}W)<0$.  In relation to entanglement
detection, if any sets $S_{\mu}$ could be subspaces rather than more complex subsets of ${\cal
H}$, then they would have a special status, as a consequence of the linearity of witness
operators. However, as we have shown above, no ${\cal S}_{\mu}$ contains the zero vector and so
none of these  sets can be subspaces of ${\cal H}$.  So, instead we shall consider the sets
\begin{equation}
\tilde{\cal S}_{\mu}={\cal S}_{\mu}{\cup}\{\tilde{0}\},
\end{equation}
where $\tilde{0}$ denotes the zero vector.  There is an important
set of circumstances under which the $\tilde{\cal S}_{\mu}$ are
indeed subspaces of ${\cal H}$. This is when the support of
$\sum_{{\mu}=1}^{M}{\rho}_{\mu}$ is the entire multiparticle
Hilbert space ${\cal H}$.  When this is the case, for any
${\mu}{\in}\{1,{\ldots},M\}$, no pure state can be orthogonal to
both ${\cal R}_{\mu}$ and ${\cal R}_{\bar{\mu}}$. So, every pure
state must be in at least one of the sets $\bar{\cal
R}^{\perp}_{\mu}$ and $\bar{\cal R}^{\perp}_{\bar{\mu}}$.  This
implies that ${\cal H}=\bar{\cal
R}_{\bar{\mu}}^{\perp}{\cup}\bar{\cal
R}_{\mu}^{\perp}{\cup}\{\tilde{0}\}$.  From this, we deduce that
\begin{eqnarray}
{\cal R}_{\bar{\mu}}^{\perp}&=&{\cal
R}_{\bar{\mu}}^{\perp}{\cap}{\cal H} \nonumber \\ &=&{\cal
R}_{\bar{\mu}}^{\perp}{\cap}\left(\bar{\cal
R}_{\bar{\mu}}^{\perp}{\cup}(\bar{\cal
R}_{\mu}^{\perp}{\cup}\{\tilde{0}\})\right) \nonumber
\\&=&({\cal
R}_{\bar{\mu}}^{\perp}{\cap}\bar{\cal
R}_{\bar{\mu}}^{\perp}){\cup}\left({\cal
R}_{\bar{\mu}}^{\perp}{\cap}(\bar{\cal
R}_{\mu}^{\perp}{\cup}\{\tilde{0}\})\right) \nonumber \\ &=&{\cal
R}_{\bar{\mu}}^{\perp}{\cap}(\bar{\cal
R}_{\mu}^{\perp}{\cup}\{\tilde{0}\})=\tilde{\cal S}_{\mu}.
\end{eqnarray}
That is, $\tilde{\cal S}_{\mu}$ is the subspace of ${\cal H}$
which is orthogonal to ${\cal R}_{\bar{\mu}}$.

Of particular interest are the projectors onto these subspaces,
which we shall denote by $P_{\mu}$.  If LOCC unambiguous state
discrimination is impossible, then at least one of the
$\tilde{\cal S}_{\mu}$ contains no product state. If these sets
nevertheless contain a non-zero vector, which is the condition for
unconstrained unambiguous state discrimination, then it follows
that all states in these $\tilde{\cal S}_{\mu}$ must be entangled.
For any entangled density operator ${\rho}^{\tilde{\cal
S}_{\mu}}_{ent}$ whose support is a subspace of such an
$\tilde{\cal S}_{\mu}$, we will have
$\mathrm{Tr}({\rho}^{\tilde{\cal S}_{\mu}}_{ent}P_{\mu})=1$.
However, for any separable density operator ${\rho}_{sep}$ with
support on ${\cal H}$, we must have
$\mathrm{Tr}({\rho}_{sep}P_{\mu})<1$. For any $\tilde{\cal
S}_{\mu}{\neq}\{\tilde{0}\}$ which contains no product states, we
can easily construct a corresponding entanglement witness which
detects the states in $\tilde{\cal S}_{\mu}$. Let
\begin{equation}
{\gamma}_{\mu}=\max_{{\rho}_{sep}}\mathrm{Tr}({\rho}_{sep}P_{\mu}).
\end{equation}
Then the operator
\begin{equation}
W_{\mu}=1_{\cal H}-{\gamma}_{\mu}^{-1}P_{\mu}
\end{equation}
is an entanglement witness which detects all of the states in
$\tilde{\cal S}_{\mu}$.

It is worth noting that, when unconstrained unambiguous state
discrimination is possible and the $S_{\mu}$ are subspaces of
${\cal H}$, then the operators $W_{\mu}$ in Eq. (16)  are always
well-defined. When there is at least one vector in $\tilde{\cal
S}_{\mu}$ besides the zero vector, then there is a product state
which is non-orthogonal to it and so both ${\gamma}_{\mu}$ and
$P_{\mu}$ are non-zero.  It follows that $W_{\mu}$ is
well-defined.  The $W_{\mu}$ can then be used to give an
alternative characterisation of the way in which the LOCC
constraint tightens further the conditions for unambiguous state
discrimination: given that unconstrained unambiguous state
discrimination is possible, for it also to be possible using an
LOCC procedure, it is necessary and sufficient that, for
${\mu}{\in}{\Delta}$, none of the $W_{\mu}$ can be entanglement
witnesses.  It is therefore possible that some of the work that
has been done on the optimisation of entanglement witnesses
\cite{Bert} could be brought to bear on the issue of LOCC
unambiguous state discimination.

It is interesting to see what happens when the states we aim to discriminate among are pure.  If
${\rho}_{\mu}=|{\psi}_{\mu}{\rangle}{\langle}{\psi}_{\mu}|$, ${\Delta}=\{1,{\ldots},M\}$ and the
$|{\psi}_{\mu}{\rangle}$ span ${\cal H}$, then the space $\tilde{\cal S}_{\mu}$ is spanned by the
reciprocal state $|{\tilde\psi}_{\mu}{\rangle}$ corresponding to $|{\psi}_{\mu}{\rangle}$
\cite{Linear,Mesymmetric}.  The existence of these states is equivalent to the linear
independence of the $|{\psi}_{\mu}{\rangle}$ and therefore their amenability to unconstrained
unambiguous state discrimination.  It follows from the theorem that if we wish to unambiguously
discriminate among them with an LOCC measurement, then the $|{\tilde\psi}_{\mu}{\rangle}$ must be
product states. Furthermore, if the $|{\psi}_{\mu}{\rangle}$ are orthonormal, then
$|{\tilde\psi}_{\mu}{\rangle}=|{\psi}_{\mu}{\rangle}$ and we obtain the result of Horodecki et al
\cite{Horodecki} that a complete set of orthonormal pure states can be unambiguously
discriminated by an LOCC measurement iff they are product states.

To summarise, we have taken advantage of the fact that any LOCC measurement is also separable and
separable measurements have a particularly simple form.  We used this to obtain a necessary
condition for LOCC unambiguous state discrimination by an LOCC, which we also proved,
constructively, to be sufficient.  For a large class of state sets. we demonstrated an
interesting connection with entanglement witnesses, in terms of which the LOCC constraint can be
explicitly expressed.

It is interesting to view this in the context of previous work relating to perfect LOCC state
discrimination \cite{Unlockable,Ghosh1,Ghosh2,Chen2,Chen3}. Here, intriguing connections with
entanglement have been revealed by many authors. The relevant issue has been entanglement
distillation.  However, it seems that unambiguous state discrimination subject to the LOCC
constraint has a particularly direct connection with entanglement detection.
\section*{Acknowledgements}

The author is grateful to the University of Hertfordshire for the
award of a Postdoctoral Research Fellowship.


\begin{thebibliography}{99}
\bibitem{support}
We take the support of a positive (semi-definite) operator to be
the subspace spanned by its eigenstates corresponding to non-zero
eigenvalues.
\end{thebibliography}
\end{document}